\documentclass[twocolumn,floatfix,showpacs,superscriptaddress,prb]{revtex4}
\usepackage{graphicx}
\usepackage{amsmath}
\usepackage{bm}
\begin{document}

\title{Valley-isospin dependence of the quantum Hall effect in a graphene \textit{p-n} junction}
\author{J. Tworzyd{\l}o}
\affiliation{Institute of Theoretical Physics, Warsaw University, Ho\.{z}a 69, 00--681 Warsaw, Poland}
\author{I. Snyman}
\affiliation{Instituut-Lorentz, Universiteit Leiden, P.O. Box 9506, 2300 RA Leiden, The Netherlands}
\author{A. R. Akhmerov}
\affiliation{Instituut-Lorentz, Universiteit Leiden, P.O. Box 9506, 2300 RA Leiden, The Netherlands}
\author{C. W. J. Beenakker}
\affiliation{Instituut-Lorentz, Universiteit Leiden, P.O. Box 9506, 2300 RA Leiden, The Netherlands}
\date{May 2007}
\begin{abstract}
We calculate the conductance $G$ of a bipolar junction in a graphene nanoribbon, in the high-magnetic field regime where the Hall conductance in the \textit{p}-doped and \textit{n}-doped regions is $2e^{2}/h$. In the absence of intervalley scattering, the result $G=(e^{2}/h)(1-\cos\Phi)$ depends only on the angle $\Phi$ between the valley isospins (= Bloch vectors representing the spinor of the valley polarization) at the two opposite edges. This plateau in the conductance versus Fermi energy is insensitive to electrostatic disorder, while it is destabilized by the dispersionless edge state which may exist at a zigzag boundary. A strain-induced vector potential shifts the conductance plateau up or down by rotating the valley isospin.
\end{abstract}
\maketitle


Recent experiments\cite{Hua07,Wil07,Ozy07} have succeeded in fabricating junctions between \textit{p}-doped and \textit{n}-doped graphene, and have begun to investigate the remarkable properties predicted theoretically.\cite{Che06,Kat06,Che07,Aba07} The conductance $G$ of a \textit{p-n} junction measures the coupling of electron-like states from the conduction band to hole-like states from the valence band, which in graphene is unusually strong because of the phenomenon of Klein tunneling.\cite{Che06,Kat06}

In the zero-magnetic field regime of Huard et al.\cite{Hua07} this coupling depends on the length scales characteristic of the \textit{p-n} interface. In the high-magnetic field regime of Williams, DiCarlo, and Marcus,\cite{Wil07} the \textit{p-n} junction has a quantized conductance, which has been explained by Abanin and Levitov\cite{Aba07} as the series conductance $G_{\rm series}=G_{p}G_{n}/(G_{p}+G_{n})$ of the quantum Hall conductances $G_{p},G_{n}$ in the \textit{p}-doped and \textit{n}-doped regions (each an odd multiple of the conductance quantum $G_{0}=2e^{2}/h$). (The \textit{p-n-p} junction experiments of \"{O}zyilmaz et al.\cite{Ozy07} are also explained in terms of a series conductance.)

These results apply if the system is sufficiently large that mesoscopic fluctuations in the conductance can be ignored, either as a consequence of self-averaging by time dependent electric fields or as a consequence of suppression of phase coherence by inelastic scattering.\cite{Aba07} In a sufficiently small system mesoscopic conductance fluctuations as a function of Fermi energy are expected to appear. In particular, in the quantum Hall effect regime, the conductance of a \textit{p-n} junction is expected to fluctuate around the series conductance $G_{\rm series}$ in a small conductor (nanoribbon) at low temperatures.

In this paper we show that a plateau in the conductance versus Fermi energy survives in the case of fully phase coherent conduction without intervalley scattering. When both \textit{p}-doped and \textit{n}-doped regions are on the lowest Hall plateau ($G_{p}=G_{n}=G_{0}$), we find a plateau at
\begin{equation}
G=\tfrac{1}{2}G_{0}(1-\cos\Phi),\label{Gphi}
\end{equation}
with $\Phi$ the angle between the valley isospins at the two edges of the nanoribbon. A random electrostatic potential is not effective at producing mesoscopic conductance fluctuations, provided that it varies slowly on the scale of the lattice constant --- so that it does not induce intervalley scattering. The dispersionless edge state that may exist at a zigzag edge (and connects the two valleys at opposite edges) is an intrinsic source of intervalley scattering when the edge crosses the \textit{p-n} interface. The angle $\Phi$ that determines the conductance plateau can be varied by straining the carbon lattice, either systematically to shift the plateau up or down, or randomly to produce a bimodal statistical distribution of the conductance in an armchair nanoribbon.

\begin{figure}[tb]
\includegraphics[width=0.7\linewidth]{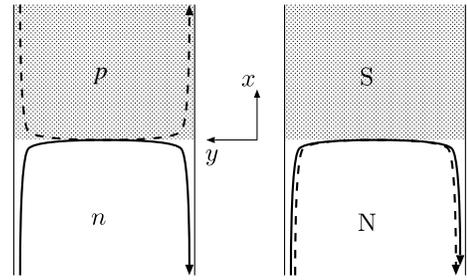}
\caption{\label{pnNS_analogy}
Schematic top view of a graphene nanoribbon containing an interface between an \textit{n}-doped and \textit{p}-doped region (left panel) and between a normal (N) and superconducting (S) region (right panel). Electron-like and hole-like edge states in the lowest Landau level are indicated by solid and dashed lines, respectively, with arrows pointing in the direction of propagation. The electron-like and hole-like valley-polarized edge states hybridize along the \textit{p-n} or NS interface to form a valley-degenerate electron-hole state. The two-terminal conductance $G=G_{0}T_{eh}$ is determined by the probability $T_{eh}$ that an electron-like state is converted into a hole-like state at the opposite edge (with $G_{0}=2e^{2}/h$ in the \textit{p-n} junction and $G_{0}=4e^{2}/h$ in the NS junction). In the absence of intervalley scattering, $T_{eh}=\frac{1}{2}(1-\cos\Phi)$, with $\Phi$ the angle between the valley isospins of the electron-like state at the two edges.\cite{Akh07}
}
\end{figure}

Our analysis was inspired by an analogy between edge channel transport of Dirac fermions along a \textit{p-n} interface\cite{Aba07} and along a normal-superconducting (NS) interface.\cite{Akh07} The analogy, explained in Fig.\ \ref{pnNS_analogy}, is instructive, but it is only a partial analogy --- as we will see. We present analytical results, obtained from the Dirac equation, as well as numerical results, obtained from a tight-binding model on a honeycomb lattice. We start with the former.

\begin{figure}[tb]
\includegraphics[width=0.7\linewidth]{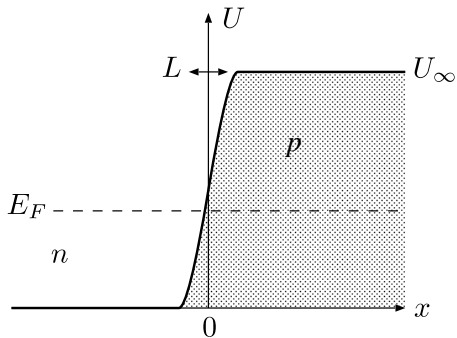}
\caption{\label{Uprofile}
Potential step at the \textit{p-n} interface (with the shaded area indicating the energy range in the valence band). The electrostatic potential $U(x)$ increases from $0$ to $U_{\infty}$ over a distance $L$ around $x=0$. The Fermi level at $E_{F}\in(0,U_{\infty})$ lies in the conduction band for negative $x$ (\textit{n}-doped region) and in the valence band for positive $x$ (\textit{p}-doped region).}
\end{figure}

The Dirac equation for massless two-dimensional fermions reads
\begin{equation}
\tau_{0}\otimes[v(\bm{p}+e{\bm A})\cdot\bm{\sigma}+U]\Psi=E\Psi,\label{Hdef}
\end{equation}
where $E$ is the energy, $v$ the Fermi velocity, $\bm{p}=(\hbar/i)(\partial/\partial x,\partial/\partial y)$ the canonical momentum operator in the $x$-$y$ plane of the graphene layer, $U(x)$ the electrostatic potential step at the \textit{p-n} interface (shown in Fig.\ \ref{Uprofile}), and $\bm{A}$ the vector potential corresponding to a perpendicular magnetic field $B$. The Pauli matrices $\sigma_{i}$ and $\tau_{i}$ act on the sublattice and valley degree of freedom, respectively (with $\sigma_{0}$ and $\tau_{0}$ representing the $2\times 2$ unit matrix). 

The Dirac equation \eqref{Hdef} is written in the valley-isotropic representation, in which the boundary condition for the wave function $\Psi$ at the edges of the nanoribbon (taken at $y=0,W$) has the form\cite{Akh07}
\begin{equation}
\Psi=(\bm{\nu}\cdot\bm{\tau})\otimes(\sin\theta\,\sigma_{x}+\cos\theta\,\sigma_{z})\Psi,\label{MIdef}
\end{equation}
parameterized by an angle $\theta$ and by the three-dimensional unit vector $\bm{\nu}$ on the Bloch sphere. The vector $\bm{\nu}$ is called the valley isospin because it represents the two-component spinor of the valley degree of freedom.\cite{note0}

An armchair edge has $\bm{\nu}\cdot\bm{\hat{z}}=0$, $\theta=\pi/2$ (modulo $\pi$), while a zigzag edge has $|\bm{\nu}\cdot\bm{\hat{z}}|=1$, $\theta=0$ (modulo $\pi$). Confinement by an infinite mass has $|\bm{\nu}\cdot\bm{\hat{z}}|=1$, $\theta=\pi/2$ (modulo $\pi$). Intermediate values of $\bm{\nu}\cdot\bm{\hat{z}}$ and $\theta$ are produced, for example, by a staggered edge potential (having a different value on the two sublattices).\cite{Son06,Akh07b} If the edge is inhomogeneous, it is the value of $\bm{\nu}$ and $\theta$ in the vicinity of the \textit{p-n} interface (within a magnetic length $l_{m}=\sqrt{\hbar/eB}$ from $x=0$) that matters for the conductance.

The boundary condition \eqref{MIdef} breaks the valley degeneracy of quantum Hall edge states,\cite{Per06,Bre06,Aba06} with different dispersion relations $E^{\pm}(q)$ for the two eigenstates $|\pm\bm{\nu}\rangle$ of $\bm{\nu}\cdot\bm{\tau}$. (We use the Landau gauge in which $\bm A$ is parallel to the boundary and vanishes at the boundary. In this gauge the canonical momentum $\hbar q$ parallel to the boundary is a good quantum number.) In the \textit{n} region (where $U=0$) the dispersion relation is determined by the equations\cite{Akh07}
\begin{subequations}
\label{Eplusminus}
\begin{align}
&f_{E^{+}}(q)=\tan(\theta/2),\;\;
f_{E^{-}}(q)=-{\rm cotan}\,(\theta/2),\label{Eplusminusa}\\
&f_{E}(q)\equiv\frac{H_{\varepsilon^{2}/2}(ql_{m})}{\varepsilon H_{\varepsilon^{2}/2-1}(ql_{m})},\;\;\varepsilon\equiv E\,l_{m}/\hbar v,\label{fdef}
\end{align}
\end{subequations}
with $H_{\alpha}(x)$ the Hermite function. The dispersion relation in the \textit{p} region is obtained by $E^{\pm}(q)\rightarrow E^{\pm}(q)+ U_{\infty}$. 

\begin{figure}[tb]
\includegraphics[width=1\linewidth]{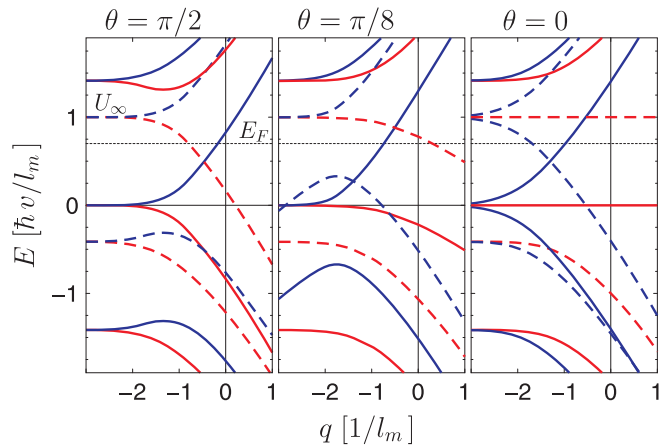}
\caption{\label{fig_dispersion1}
Dispersion relation $E^{\pm}(q)$ according to Eq.\ \eqref{Eplusminus} of edge states near the Dirac point in the \textit{n} region (solid curves) and in the \textit{p} region (dashed curves). The color of the curves indicates the valley polarization (blue: $+\bm{\nu}$, red; $-\bm{\nu}$). The three panels correspond to three different boundary conditions, and illustrate the transition from an armchair edge (leftmost panel) to a zigzag edge (rightmost panel).
}
\end{figure}

The dispersion relation near the Dirac point ($E=0$) is plotted in Fig.\ \ref{fig_dispersion1} for three values of $\theta$. (It does not depend on $\bm{\nu}$.) For any $\theta\neq 0$ (modulo $\pi$) there is a nonzero interval $\Delta E_{F}$ of Fermi energies in which just two edge channels of {\em opposite\/} valley isospin cross the Fermi level (dotted line), one electron-like edge channel from the \textit{n} region (blue solid curve) and one hole-like edge channel from the \textit{p} region (red dashed curve). The case $\theta=0$ is special because of the dispersionless edge state which extends along a zigzag boundary.\cite{Fuj96} As $\theta\rightarrow 0$ the interval $\Delta E_{F}$ shrinks to zero, and at $\theta=0$ (modulo $\pi$) the electron-like and hole-like edge channels in the lowest Landau level have {\em identical\/} valley isospins. It is here that the analogy with the problem of the NS junction\cite{Akh07} stops, because in that problem the electron and hole edge channels at the Fermi level have opposite valley isospins irrespective of $\theta$.

\begin{figure}[tb]
\includegraphics[width=0.7\linewidth]{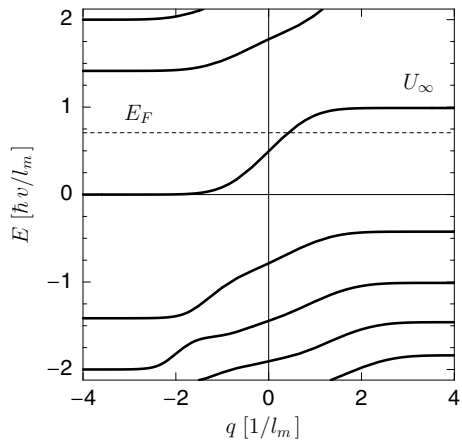}
\caption{\label{fig_dispersion2}
Dispersion relation at the \textit{p-n} interface, calculated numerically from the Dirac equation for a step function potential profile. Each Landau level has a twofold valley degeneracy.
}
\end{figure}

The two valley-polarized edge channels from the \textit{n} and \textit{p} regions are coupled by the potential step at the \textit{p-n} interface. Edge states along a potential step which is smooth on the scale of the lattice constant $a$ are valley degenerate,\cite{Luk07,Mil07} because an electrostatic potential in the Dirac equation does not couple the valleys. The dispersion relation, for the case of an abrupt potential step ($a\ll L\ll l_{m}$), is plotted in Fig.\ \ref{fig_dispersion2}. (It is qualitatively similar for $L\gg l_{m}$.) The Fermi level now intersects with a two-fold valley degenerate edge channel of mixed electron-hole character.

The two-terminal conductance of the \textit{p-n} junction is given by\cite{Aba07} $G=G_{0}T_{eh}$, in terms of the probability $T_{eh}$ that an electron incident in an electron-like edge channel along the left edge is transmitted to a hole-like edge channel along the right edge. We now show that this probability takes on a universal form, dependent only on the valley isospins at the edge, in the absence of intervalley scattering. The argument is analogous to that in the NS junction,\cite{Akh07} and requires that the electron-like and hole-like edge channels at the same edge have opposite valley isospins ($\pm\bm{\nu}_{L}$ for the left edge and $\pm\bm{\nu}_{R}$ for the right edge).\cite{note1}

Since the unidirectional motion of the edge states prevents reflections, the total transmission matrix $t_{\rm total}=t_{R}t_{pn}t_{L}$ from one edge to the other edge is the product of  three $2\times 2$ unitary matrices: the transmission matrix $t_{L}$ from the left edge to the \textit{p-n} interface, the transmission matrix $t_{pn}$ along the interface, and the transmission matrix $t_{R}$ from the \textit{p-n} interface to the right edge. In the absence of intervalley scattering $t_{pn}=e^{i\phi_{pn}}\tau_{0}$ is proportional to the unit matrix, while
\begin{equation}
t_{X}=e^{i\phi_{X}}|+\bm{\nu}_{X}\rangle\langle+\bm{\nu}_{X}|+e^{i\phi'_{X}}|-\bm{\nu}_{X}\rangle\langle-\bm{\nu}_{X}|\label{tpndef}
\end{equation}
(with $X=L,R$) is diagonal in the basis $|\pm\bm{\nu}_{X}\rangle$ of eigenstates of $\bm{\nu}_{X}\cdot\bm{\tau}$. The phase shifts $\phi_{pn},\phi_{X},\phi'_{X}$ need not be determined. Evaluation of the transmission probability 
\begin{equation}
T_{eh}=|\langle+\bm{\nu}_{L}|t_{\rm total}|-\bm{\nu}_{R}\rangle|^{2}\label{Tehdef}
\end{equation}
leads to the conductance \eqref{Gphi} with $\cos\Phi=\bm{\nu}_{L}\cdot\bm{\nu}_{R}$.


\begin{figure}[tb]
\includegraphics[width=0.9\linewidth]{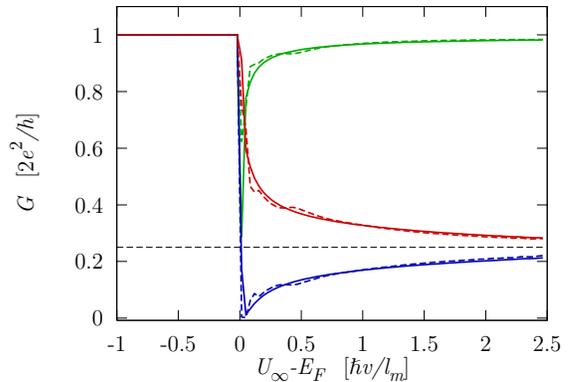}
\caption{\label{fig_disorder1}
Conductance of an armchair nanoribbon containing the potential step $U(x)=\frac{1}{2}[\tanh(2x/L)+1]U_{\infty}$, calculated numerically from the tight-binding model in a perpendicular magnetic field ($l_{m}=5\,a$). The step height $U_{\infty}$ is varied from below $E_{F}$ (unipolar regime) to above $E_{F}$ (bipolar regime), at fixed $E_{F}=\hbar v/l_{m}$ and $L=50\,a$. The solid curves are without disorder, while the dashed curves are for a random electrostatic potential landscape ($K_{0}=1$, $\xi=10\,a$). The number ${\cal N}$ of hexagons across the ribbon is 97 (red curves), 98 (blue), and 99 (green). The dashed horizontal line marks the plateau at $G=\frac{1}{4}\times 2e^{2}/h$.
}
\end{figure}

\begin{figure}[tb]
\includegraphics[width=0.9\linewidth]{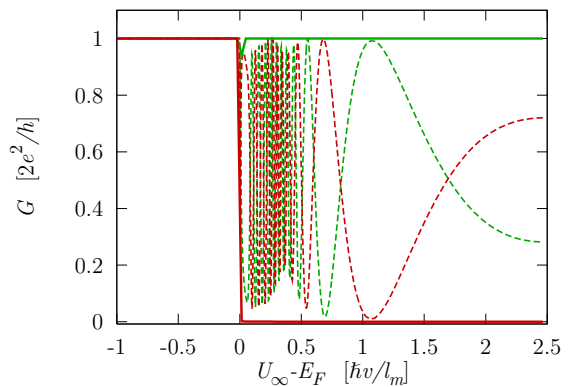}
\caption{\label{fig_disorder2}
Same as Fig.\ \ref{fig_disorder1}, for the case of a zigzag nanoribbon (${\cal N}=114$ for the green curves and $115$ for the red curves).
}
\end{figure}

To test the robustness of the conductance plateau to a random electrostatic potential, we have performed numerical simulations. A random potential landscape is introduced in the same way as in Ref.\ \onlinecite{Ryc06}, by randomly placing impurities at $N_{\rm imp}$ sites $\bm{R}_{i}$ on a honeycomb lattice. Each impurity has a Gaussian potential profile $U_{i}\exp(-|\bm{r}-\bm{R}_{i}|^{2}/2\xi^{2})$ of range $\xi$ and random height $U_{i}\in(-\delta,\delta)$. We take $\xi$ equal to the mean separation $d$ of the impurities and large compared to the lattice constant $a$. The strength of the resulting potential fluctuations $\delta U(\bm{r})$ is quantified by the dimensionless correlator
\begin{equation}
K_{0}=\frac{{\cal A}}{(\hbar v)^{2}}\frac{1}{N_{\rm tot}^{2}}\sum_{i,j=1}^{N_{\rm tot}}\langle \delta U(\bm{r}_{i})\delta U(\bm{r}_{j})\rangle,\label{K0def}
\end{equation}
where the sum runs over all $N_{\rm tot}$ lattice sites $\bm{r}_{i}$ in a nanoribbon of area ${\cal A}$. 

Results are shown in Figs.\ \ref{fig_disorder1} and \ref{fig_disorder2} for an armchair and zigzag nanoribbon, respectively. The angle $\Phi$ between the valley isospins at two opposite armchair edges depends on the number ${\cal N}$ of hexagons across the ribbon: $\Phi=\pi$ if ${\cal N}$ is a multiple of $3$, $|\Phi|=\pi/3$ if it is not.\cite{Bre06b} Fig.\ \ref{fig_disorder1} indeed shows that the conductance as a function of $U_{\infty}-E_{F}$ switches from a plateau at the $\Phi$-independent Hall conductance $G_{0}$ in the unipolar regime ($U_{\infty}<E_{F}$) to a $\Phi$-dependent value given by Eq.\ \eqref{Gphi} in the bipolar regime ($U_{\infty}>E_{F}$). The plateau persists in the presence of a smooth random potential (compare solid and dashed curves in Fig.\ \ref{fig_disorder1}). By reducing the potential range we found that the plateaus did not disappear until $\xi\lesssim 3\,a$ (not shown). 

As expected in view of the intervalley scattering produced by the dispersionless edge state in a zigzag nanoribbon, no such robust conductance plateau exists in this case (Fig.\ \ref{fig_disorder2}). In the presence of disorder the conductance oscillates around its ensemble average $G_{0}/2$, in a sample specific manner. The numerics for any given realization of the disorder potential satisfies approximately the sum rule $G({\cal N})+G({\cal N}+1)\approx G_{0}$, for which we have not yet found an analytical derivation.


The valley-isospin dependence of the quantum Hall effect in a \textit{p-n} junction makes it possible to use {\em strain\/} as a means of variation of the height of the conductance plateaus. Strain introduces a vector potential term $ev\tau_{z}\otimes(\delta\bm{A}\cdot\bm{\sigma})\Psi$ in the Dirac equation \eqref{Hdef}, corresponding to a fictitious magnetic field of opposite sign in the two valleys.\cite{Kan97,Suz02,Mor06,Mor06b} This term rotates the Bloch vector of the valley isospin around the $z$-axis, which in the case of an armchair nanoribbon corresponds to a rotation of the valley isospin in the $x$-$y$ plane. Strain may appear locally at an armchair edge by passivation of the carbon bonds.\cite{Son06} (The resulting change $\delta\tau$ of the hopping energy $\tau$ changes $\Phi$ by an amount\cite{Nov07} $\delta\Phi=2\sqrt{3}\,\delta\tau/\tau$.) Random strain along the \textit{p-n} interface, resulting from mesoscopic corrugation of the carbon monolayer,\cite{Mor06b} corresponds to a random value of the angle $\Phi$ in the conductance formula \eqref{Gphi}. A uniform distribution of $\Phi$ implies a bimodal statistical distribution of the conductance,
\begin{align}
P(G)&=\frac{1}{\pi}\int_{0}^{\pi}d\Phi\,\delta\bigl[G-\tfrac{1}{2}G_{0}(1-\cos\Phi)\bigr]\nonumber\\
&= [\pi^{2}G(G_{0}-G)]^{-1/2},\;\;0<G<G_{0},\label{PGresult}
\end{align}
distinct from the uniform distribution expected for random edge channel mixing.\cite{Aba07}

In summary, we have presented analytical and numerical evidence for the existence of a valley-isospin dependent conductance plateau in a \textit{p-n} junction in the quantum Hall effect regime. In recent experiments\cite{Wil07,Ozy07} the conductance was simply the series conductance of the \textit{p}-doped and \textit{n}-doped regions, presumably because of local equilibration. We have shown that the mesoscopic fluctuations, expected to appear in the phase coherent regime,\cite{Aba07} are suppressed in the absence of intervalley scattering. The conductance plateau is then not given by the series conductance, but by Eq.\ \eqref{Gphi}. The same formula applies to the conductance of a normal-superconducting junction in graphene,\cite{Akh07} revealing an intriguing analogy between Klein tunneling in \textit{p-n} junctions and Andreev reflection at NS interfaces.\cite{Bee06,note2}


This research was supported by the Dutch Science Foundation NWO/FOM  and by the European Community's Marie Curie Research Training Network (contract MRTN-CT-2003-504574, Fundamentals of Nanoelectronics).


\begin{thebibliography}{99}
\bibitem{note0}
The two eigenstates $|+\bm{\nu}\rangle$ and $|-\bm{\nu}\rangle$ of $\bm{\nu}\cdot\bm{\tau}$ (defined by  $\bm{\nu}\cdot\bm{\tau}|\pm\bm{\nu}\rangle=\pm|\pm\bm{\nu}\rangle$) are states of definite valley polarization (parallel or antiparallel to the unit vector $\bm{\nu}$). When the valley isospin $\bm{\nu}$ points to the north pole or south pole on the Bloch sphere, the polarization is such that the eigenstate lies entirely within one single valley. This is the case for the zigzag edge or for the infinite mass confinement (for which $|\bm{\nu}\cdot\bm{\hat{z}}|=1$). When the valley isospin points to the equator, the eigenstate is a coherent equal-weight superposition of the two valleys. This is the case for the armchair edge (for which $\bm{\nu}\cdot\bm{\hat{z}}=0$).
\bibitem{note1} Eq.\ \eqref{MIdef} is invariant under the transformation $\bm{\nu}\rightarrow -\bm{\nu}$, $\theta\rightarrow\theta+\pi$. This freedom is used to choose the sign of $\bm{\nu}$ such that the electron-like edge channel has isospin $+\bm{\nu}$.
\bibitem{note2}  With hindsight, the same analogy can be noted between the phenomena of negative refraction of Ref.\ \onlinecite{Che07} and Andreev retroreflection of Ref.\ \onlinecite{Bee06}.
\end{thebibliography}
\end{document}